\documentclass[showpacs,preprintnumbers]{revtex4}
\usepackage{graphicx,amsmath,amssymb,color}
\begin{document}
\title{Superfluidity of Dipolar Excitons in a Black
Phosphorene Double Layer}
\author{ Oleg L. Berman$^{1,2}$,  Godfrey Gumbs$^{2,3,4}$, and Roman Ya. Kezerashvili$^{1,2}$}
\affiliation{$^{1}$Physics Department,  New York City College of Technology\\
The City University of New York,
  300 Jay Street,   Brooklyn NY, 11201, USA \\
$^{2}$The Graduate School and University Center\\
The City University of New York,
New York, NY 10016, USA \\
$^{3}$Department of Physics and Astronomy, Hunter College of the
City University of New York,
695 Park Avenue, New York, NY 10065\\
$^{4}$Donastia International Physics Center (DIPC),
P de Manuel Lardizabal, 4, 20018 San Sebastian, Basque Country, Spain\\
 }
\date{\today }

\begin{abstract}
We study the formation of dipolar excitons and their superfluidity
in a  black phosphorene double layer. The analytical expressions for
the single dipolar exciton energy spectrum and wave function are
obtained. It is predicted that a weakly interacting gas of dipolar
excitons in a double layer of black phosphorus exhibits
superfluidity due to the dipole-dipole repulsion between the dipolar
excitons. In calculations are employed the Keldysh and Coulomb
potentials for the interaction between the charge carriers to
analyze the influence of the screening effects on the studied
phenomena. It is shown that the critical velocity of superfluidity,
the spectrum of collective excitations, concentrations of the
superfluid and normal component, and mean field critical temperature
for superfluidity are anisotropic and demonstrate the dependence on
the direction of motion of dipolar excitons. The critical
temperature for superfluidity increases if the exciton concentration
and the interlayer separation increase. It is shown that the dipolar
exciton binding energy and mean field critical temperature for
superfluidity are sensitive to the electron and hole effective
masses. The proposed experiment to observe a directional
superfluidity of excitons is addressed.

\end{abstract}


\maketitle

\section{Introduction}
\label{sec1}

The Bose-Einstein condensation (BEC) and superfluidity of dipolar
(indirect) excitons, formed by  electrons and holes, spatially
separated in two parallel two-dimensional (2D) layers of
semiconductor, were proposed~\cite{Lozovik} and recent progress on
BEC of semiconductor dipolar excitons was
reviewed~\cite{Moskalenko_Snoke,Combescot}. Due to relatively large
exciton binding energies in novel 2D semiconductors, the BEC and
superfluidity of dipolar excitons in double layers of transition
metal dichalcogenides (TMDCs) was
studied~\cite{Fogler,MacDonald,BK}.

\medskip
\par

Phosphorene, an atom-thick layer of the black phosphorus
\cite{Warren_ACS} that does have a natural band gap, has aroused
considerable interest currently. It has been shown that monolayer
black phosphorene is an relatively unexplored two dimensional
semiconductor with a high hole mobility and exhibits unique
many-electron effects~\cite{1a}. In particular, first principles
calculations have predicted unusual strong anisotropy for the
in-plane thermal conductivity  in these materials~\cite{1}. Among
the intriguing band structure features found are large excitonic
binding energy \cite{LiuACS2014,Tran2014}, prominent anisotropic
electron and hole effective masses
\cite{Natute2014,Appelbaum,Rodin,Chaves2015} and carrier mobility
\cite{Xia2014,Natute2014}.  Recently the exciton binding energy for
direct excitons in monolayer black phosphorus, placed on a
SiO$_\mathrm{2}$ substrate was obtained experimentally by
polarization-resolved photoluminescence measurements at room
temperature \cite{5b}. External perpendicular electric
fields~\cite{3} and mechanical strain~\cite{4,4a} have been applied
to demonstrate that  the electronic properties of phosphorene  may
be significantly modified.  According to
Refs.~[\onlinecite{Tran2014,5b}], excitons and highly anisotropic
optical responses of few-layer black phosphorous  may be possible.
Specifically,
 black phosphorous absorbs light polarized along its armchair direction and is
transparent to light polarized along the zigzag direction.  Consequently,
black phosphorene  may be employed as a viable linear polarizers. Also the interest in these recently fabricated
2D phosphorene crystals has  been growing  because
they have displayed potential for applications in electronics
 including field effect transistors~\cite{2}.

\medskip
\par

This paper explores the way in which the anisotropy of black
phosphorene is capable  of affecting superfluidity in double layer
structure. While it is important to mention that whereas the exciton
binding energy was calculated using density functional theory (DFT)
and quasiparticle self-consistent GW methods for direct excitons in
suspended few-layer black phosphorus~\cite{Tran2014}, here we apply
an analytical approach for indirect excitons in a phosphorene double
layer. In our model, electrons and holes are confined to two
separated parallel phosphorene layers which are embedded in a
dielectric medium. We have taken screening of the interaction
potential between an electron and hole through the Keldysh potential
~\cite{Keldysh}. The dilute system of dipolar excitons form a weakly
interacting Bose gas, which can can be treated in the Bogoliubov
approximation~\cite{Abrikosov}. The anisotropic dispersion relation
for the single dipolar exciton in a phosphorene double layer results
in the angle dependent spectrum of collective excitations with the
angle dependent sound velocity, which causes the dependence of the
critical velocity for the superfluidity on the direction of motion
of dipolar excitons.
 While the concentrations of the normal and superfluid components
for the BCS-type fermionic superfluid with the anisotropic order
parameter do not depend on the direction of motion of the Cooper
pairs~\cite{Saslow}, we obtain the concentrations of the normal and
superfluid components for dipolar excitons in a double layer
phosphorene to be dependent on the directions of motion of excitons.
Therefore, the mean field temperature of the superfluidity for
dipolar excitons in a phosphorene double layer also depends on the
direction of motion of the dipolar excitons. At some fixed
temperatures, the motion of dipolar excitons in some directions is
superfluid, while in other directions is dissipative. This effect
makes superfluidity of dipolar excitons in a phosphorene double
layer to be different from other 2D semiconductors, due to high
anisotropy of the dispersion relations for the charge carriers in
phosphorene.    The calculations have been performed for both the
Keldysh and Coulomb potentials, describing the interactions between
the charge carriers. Such approach allows to analyze the influence
of the screening effects on the properties of a weakly interacting
Bose gas of dipolar excitons in a phosphorene double layer. We also
study the dependence of the binding energy, the sound velocity, and
the mean field temperature of the superfluidity for dipolar excitons
on the electron and hole effective masses.

\medskip
\par

The paper is organized in the following way.  In Sec.~\ref{tm}, the
energy spectrum and wave functions  for a single dipolar exciton in
a phosphorene double layer are obtained, and the dipolar exciton
effective masses and binding energies are calculated. The angle
dependent spectrum of collective excitations and the sound velocity
for the dilute weakly interacting Bose gas of dipolar excitons in
the Bogoliubov approximation are derived in Sec.~\ref{collect}. In
Sec.~\ref{super}, the concentrations of the normal and superfluid
components and the mean field critical temperature of superfluidity
 are obtained. The proposed experiment to
study the superfluidity of dipolar excitons in  different directions
of motion of dipolar excitons is  discussed in
Sec.~\ref{experiment}. The conclusions follow in Sec.~\ref{conc}.

\section{Theoretical Model}

\label{tm}

\medskip
\par

\begin{figure}[h]
\includegraphics[width=14.0cm]{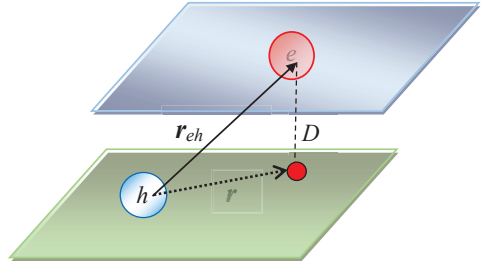} \vspace{-12cm}
\caption{(Color online) Schematic illustration of a dipolar exciton
consisting of a spatially separated electron and hole in a black
phosphorene double layer.} \label{Fig1}
\end{figure}

In the system under consideration in this paper, electrons are
 confined in a 2D phosphorene monolayer, while an
equal number of positive holes are located in a parallel phosphorene
monolayer at a distance $D$ away. The system of the
charge carriers in two parallel phosphorene layers is treated as a
two-dimensional system without interlayer hopping. In this system,
the electron-hole recombination due to the tunneling of electrons
and holes between different phosphorene monolayers is suppressed by
the dielectric barrier with the dielectric constant $\varepsilon
_{d}$ that separates the phosphorene monolayers. Therefore, the
dipolar excitons, formed by electrons and holes, located in two
different phosphorene monolayers, have a longer lifetime than the
direct excitons. The electron and hole via electromagnetic interaction $%
V(r_{eh}),$ where $r_{eh}$ is distance between the electron and hole, could form a bound state, i.e.,
an exciton, in three-dimensional (3D) space. Therefore, to
determine the binding energy of the exciton one must solve a two
body problem in restricted 3D space.  However, if one projects the
electron position vector onto the black phosphorene plane with holes
and replace the relative coordinate vector ${\bf r}_{eh}$ by its
projection $\mathbf{r}$ on this plane, the potential $V(r_{eh})$ may
be expressed as $V(r)= V(\sqrt{r^{2}+D^{2}}),$ where $r$ is the
relative distance between the hole and the projection of the
electron position vector onto the phosphorene plane with holes. A
schematic illustration of the exciton is presented in Fig.
\ref{Fig1}. By introducing in-plane coordinates
$\mathbf{r}_{1}=(x_{1},y_{1})$ and $\mathbf{r}_{2}=(x_{2},y_{2})$\
for the electron and the projection vector of the hole,
respectively, so that $\mathbf{r}=$ $\mathbf{r}_{1}-\mathbf{r}_{2}$,
one can describe the exciton by employing a two-body 2D
Schr\"{o}dinger equation with potential $V(\sqrt{r^{2}+D^{2}}).$ In
this way, we have reduced the restricted 3D two-body problem to a 2D
two-body problem on a phosphorene layer with the holes.

\subsection{Hamiltonian for an electron-hole pair in a black phosphorene double layer}
\label{single}

\medskip
\par

 Within the framework of our model the
coordinate vectors of the electron and hole may be replaced by their
2D projections onto the plane of one phosphorene layer. These
in-plane coordinates $\mathbf{r}_{1}=(x_{1},y_{1})$ and
$\mathbf{r}_{2}=(x_{2},y_{2})$ for an electron and a hole,
respectively, will be used in our  description. We assume that at
low momentum $\mathbf{p}=(p_{x},p_{y})$,  i.e.,  near the $\Gamma$
point, the single electron and hole energy spectrum
$\varepsilon _{l}^{(0)}(\mathbf{p})$ is given by

\begin{eqnarray}
\varepsilon_{l}^{(0)}(\mathbf{p}) = \frac{p_{x}^{2}}{2m_{x}^{l}} +
\frac{ p_{y}^{2}}{2m_{y}^{l}}, \ \ l =e,\ h, \label{esingl}
\end{eqnarray}
where $m_{x}^{l}$ and $m_{y}^{l}$ are the electron/hole effective
masses along the $x$ and $y$ directions, respectively.   We assume
that $OX$ and $OY$ axes correspond to the armchair and zigzag
directions in a phosphorene monolayer, respectively.

\medskip
\par

 The model  Hamiltonian within the effective mass approximation for
a single electron-hole pair in a black phosphorene
double layer is given by

\begin{eqnarray}
\hat{H}_{0} = -
\frac{\hbar^{2}}{2m_{x}^{e}}\frac{\partial^{2}}{\partial x_{1}^{2}}
+ \frac{\hbar^{2}}{2m_{y}^{e}}\frac{\partial^{2}}{\partial
y_{1}^{2}}  -
\frac{\hbar^{2}}{2m_{x}^{h}}\frac{\partial^{2}}{\partial x_{2}^{2}}
- \frac{\hbar^{2}}{2m_{y}^{h}}\frac{\partial^{2}}{\partial
y_{2}^{2}} + V\left(\sqrt{r^{2}+D^{2}}\right)\ , \label{H0}
\end{eqnarray}
where $V\left(\sqrt{r^{2}+D^{2}}\right)$ is the potential energy for
electron-hole pair attraction, when the electron and hole are
located in two different 2D planes. To separate the relative motion
of the electron-hole pair from their center-of-mass motion one can
 introduces  variables for the  center-of-mass of an
electron-hole pair $\mathbf{R} =(X,Y)$ and the relative motion of an electron and a hole $%
\mathbf{r} = (x,y)$, as $X =  (m_{x}^{e} x_{1} + m_{x}^{h} x_{2})/(
m_{x}^{e}+ m_{x}^{h})$, \
 $Y = (m_{x}^{e} y_{1} + m_{x}^{h} y_{2})/(
m_{x}^{e}+ m_{x}^{h})$,  \   $x = x_{1} - x_{2} \ ,
 y = y_{1} - y_{2}$ \ ,   $r^{2}=x^2+y^2$.
The latter allows to rewrite the Hamiltonian as $\hat{H}_{0} =
\hat{H}_{c} + \hat{H}_{rel} $,
 where
\begin{eqnarray}
\hat{H}_{c} = - \frac{\hbar^{2}}{2M_{x}}\frac{\partial^{2}}{\partial X^{2}}
- \frac{\hbar^{2}}{2M_{y}}\frac{\partial^{2}}{\partial Y^{2}} \ ,
\label{Hc}
\end{eqnarray}
\begin{eqnarray}
\hat{H}_{rel} = - \frac{\hbar^{2}}{2\mu_{x}}\frac{\partial^{2}}{\partial
x^{2}} - \frac{\hbar^{2}}{2\mu_{y}}\frac{\partial^{2}}{\partial y^{2}} + V(%
\sqrt{r^{2}+D^{2}}) \
\label{Hrel}
\end{eqnarray}
are the Hamiltonians of the center-of-mass and relative motion of an
electron-hole pair, respectively. In Eqs.\ (\ref{Hc}) and
(\ref{Hrel}), $M_{x} = m_{x}^{e}+ m_{x}^{h}$ and $M_{y} = m_{y}^{e}+
m_{y}^{h}$ are the effective exciton masses, describing the motion
of an electron-hole center-of-mass in the  $x$ and $y$
directions, respectively, while $\mu_{x} = \frac{m_{x}^{e} m_{x}^{h}}{%
m_{x}^{e}+ m_{x}^{h}}$ and $\mu_{y} = \frac{m_{y}^{e} m_{y}^{h}}{%
m_{y}^{e}+ m_{y}^{h}}$ are the reduced masses, describing the
relative motion of an electron-hole pair in the $x$ and $y$
directions, respectively.

\medskip
\par

In general the Schr\"{o}dinger equation for this electron-hole pair has the form:
$\hat{H}_{0} \Psi(\mathbf{r}_{1},\mathbf{r}_{2}) = E\Psi(\mathbf{r}_{1},%
\mathbf{r}_{2})$, where $\Psi(\mathbf{r}_{1},\mathbf{r}_{2})$ and
$E$ are its eigenfunction and eigenenergy. Substituting      Eqs.\
(\ref{Hc}) and (\ref{Hrel}) into $\hat{H}_{0} $,  due to the
separation of variables
for the center-of-mass and relative motion, one can write $\Psi(\mathbf{r}_{1},%
\mathbf{r}_{2})$ in the form $\Psi(\mathbf{r}_{1},\mathbf{r}_{2}) = \Psi(%
\mathbf{R},\mathbf{r}) = e^{i\mathbf{P}\cdot\mathbf{R}/\hbar}\varphi(\mathbf{%
r})$, where $\mathbf{P} = (P_{x},P_{y})$ is the momentum for the
center-of-mass of the electron-hole pair and $\varphi(\mathbf{r})$
is the wave function for the electron-hole pair, given by the 2D
Schr\"{o}dinger equation:

\begin{eqnarray}
\left[ - \frac{\hbar^{2}}{2\mu_{x}}\frac{\partial^{2}}{\partial x^{2}} -
\frac{\hbar^{2}}{2\mu_{y}}\frac{\partial^{2}}{\partial y^{2}} + V\left(\sqrt{
r^{2}+D^{2}}\right)\right]\varphi(x,y) = \mathcal{E}\varphi(x,y) ,
\label{Schrel}
\end{eqnarray}
where $\mathcal{E}$ is the eigenenergy
 of  the electron-hole pair in a black phosphorene double layer.

\subsection{Electron-hole interaction in a black phosphorene double layer}

\medskip
\par

The electromagnetic interaction in a thin layer of material has a
nontrivial form due to screening \cite{Keldysh,Rubio}. Whereas the
electron and hole are interacting via the Coulomb potential, in
black phosphorene the electron-hole interaction is affected by
screening which causes the electron-hole attraction to be described
by the Keldysh potential \cite{Keldysh}. This potential has been
widely used to describe the electron-hole interaction in TMDC
\cite{Reichman,Prada,Saxena,VargaPRB2016,Kezerashvili2016} and black phosphorene \cite{Rodin,Chaves2015,Katsnelson} monolayers.
The Keldysh potential has the form~\cite{Rodin}

\begin{eqnarray}
V(r_{eh}) = -\frac{\pi k e^{2}}{\left(\varepsilon_{1} +
\varepsilon_{2}\right)\rho_{0}} \left[H_{0}\left(\frac{r_{eh}}{\rho_{0}}%
\right) - Y_{0}\left(\frac{r_{eh}}{\rho_{0}}\right) \right] ,
\label{Keldysh}
\end{eqnarray}
where $r_{eh}$ is the distance between the
electron and hole located in the different parallel planes, $k=9\times 10^{9}\ N\times m^{2}/C^{2}$,
$H_{0}(x)$ and $Y_{0}(x)$ are Struve and Bessel functions of the
second kind of order $\nu=0$, respectively, $\varepsilon_{1}$ and $\varepsilon_{2}$
denote the background dielectric constants of the dielectrics,
surrounding the black phosphorene layer, and the screening length
$\rho_{0}$ is defined by $\rho _{0}=2\pi
\zeta/\left[\left(\varepsilon_{1} +
\varepsilon_{2}\right)/2\right]$, where $\zeta = 4.1 \
{\AA}$~\cite{Rodin}. Assuming that the dielectric between two
phosphorene monolayers is the same as substrate material with
dielectric constant $\varepsilon_{d}$, we set $\varepsilon_{1} =
\varepsilon_{2} = \varepsilon_{d}$. The screening length $\rho_{0}$
determines the boundary between two different behaviors for the
potential due to a nonlocal macroscopic screening. For large
separation between the electron and hole, i.e., $r_{eh} \gg \rho
_{0\text{ }}$, the potential has the three-dimensional Coulomb tail.
On the other hand, for small $r_{eh} \ll \rho _{0\text{ }}$distances
it becomes a logarithmic Coulomb potential of interaction between
two point charges in 2D. \ A crossover between these two regimes
takes place around distance $\rho _{0}$.

\medskip
\par

Making use of  $r_{eh} = \sqrt{r^{2}+ D^{2}}$ in Eq.~(\ref{Keldysh})
and assuming that $r\ll D$, one can expand Eq.\ (\ref{Keldysh}) as a
Taylor series in terms of $\left(r/D\right)^{2}$. By limiting
ourselves to the first order with respect to $\left(r/D\right)^{2}$,
we obtain

\begin{eqnarray}
V(r) = -V_{0} + \gamma r^{2}, \label{expand}
\end{eqnarray}
with

\begin{eqnarray}
V_{0} &=& \frac{\pi k e^{2}}{\left(\varepsilon_{1} +
\varepsilon_{2}\right)\rho_{0}} \left[H_{0}\left(\frac{D}{\rho_{0}}\right) -
Y_{0}\left(\frac{D}{\rho_{0}}\right) \right],  \label{V0}
 \nonumber\\
\gamma &=& - \frac{\pi k e^{2}}{2\left(\varepsilon_{1} +
\varepsilon_{2}\right)\rho_{0}^{2}D} \left[H_{-1}\left(\frac{D}{\rho_{0}}%
\right) - Y_{-1}\left(\frac{D}{\rho_{0}}\right) \right] ,
\label{gamma}
\end{eqnarray}
where $H_{-1}\left(\frac{D}{\rho_{0}}\right)$ and $Y_{-1}\left(\frac{D}{%
\rho_{0}}\right)$ are Struve and  Bessel functions of the
second kind of order $\nu=-1$, respectively.

\medskip
\par

To illustrate the screening effect of the Keldysh interaction let us
use for the electron-hole interaction the Coulomb potential. The
potential energy of the electron-hole attraction in this case is
$V(r)=-ke^{2}/(\epsilon _{d}\sqrt{r^{2}+D^{2}})$. Assuming $r\ll D$
and retaining only the first two terms of the Taylor series, one
obtains the same form for a potential as  Eq.~(\ref{expand})  but
with the following expressions for $V_{0}$ and $\gamma$:

\begin{eqnarray}
V_{0}=\frac{ke^{2}}{\epsilon _{d}D},\hspace{1cm}\gamma =\frac{ke^{2}}{%
2\epsilon _{d}D^{3}}.
\label{V0g}
\end{eqnarray}

Replacement of $V\left(\sqrt{r^{2}+D^{2}}\right)$ in Eq.\  (\ref{Hrel}) by the potential (\ref{expand}) allows to reduce the problem of
indirect exciton formed between two layers to an exactly solvable two-body problem as this is demonstrated in the next subsection.
\medskip
\par

\subsection{Wave function and binding energy of an exciton}

Substituting (\ref{expand}) with parameters (\ref{gamma}) for the Keldysh potential or
(\ref{V0g}) for the Coulomb potential, into Eq.\  (\ref{Hrel}) and
using  $r^2=x^2+y^2$, one obtains an equation which has the form of
the Schr\"{o}dinger equation for a 2D anisotropic harmonic
oscillator. This equation allows to separate the $x$ and $y$
variables and can be reduced to two independent Schr\"{o}dinger
equations for 1D harmonic oscillators, i.e.,

\begin{eqnarray}
&-& \frac{\hbar^{2}}{2\mu_{x}}\frac{d^{2}}{d x^{2}}\psi(x) + \gamma
x^{2}\psi (x) = \left( \mathcal{E}_{x}+\frac{V_{0}}{2}\right)\psi
(x),  \nonumber \\
&-& \frac{\hbar^{2}}{2\mu_{y}}\frac{d^{2}}{d y^{2}}\psi(y) + \gamma
y^{2}\psi (y) = \left( \mathcal{E}_{y}+\frac{V_{0}}{2}\right)
\psi(y), \label{1DSchrel}
\end{eqnarray}
which have eigenfunctions given by \cite{Landau}:

\begin{eqnarray}
\psi_{n} (x) &=& \frac{1}{\pi^{1/4}a_{x}^{1/2}}\frac{1}{\sqrt{2^{n}n!}}
e^{-x^{2}/\left(2a_{x}^{2}\right)}\mathcal{H}_{n}\left(\frac{x}{a_{x}}%
\right),  \nonumber \\
\psi_{m} (y) &=& \frac{1}{\pi^{1/4}a_{y}^{1/2}}\frac{1}{\sqrt{2^{m}m!}}
e^{-y^{2}/\left(2a_{y}^{2}\right)}\mathcal{H}_{m}\left(\frac{y}{a_{y}}%
\right),
\label{1dapsu}
\end{eqnarray}
where $n=0,1,2,3,\ldots$ and $m=0,1,2,3,\ldots$ are the quantum numbers, $%
\mathcal{H}_{n}(\xi)$ are Hermite polynomials, and $a_{x}=\left( \hbar /%
\sqrt{2\mu_{x}\gamma }\right)^{1/2}$ and $a_{y}=\left( \hbar /\sqrt{%
2\mu_{y}\gamma }\right)^{1/2}$, respectively. The corresponding
eigenenergies for the 1D harmonic oscillators
 are given by~\cite{Landau}:

\begin{eqnarray}
\mathcal{E}_{xn} &=& - \frac{V_{0}}{2} + \hbar\sqrt{\frac{2\gamma}{\mu_{x}}}%
\left(n + \frac{1}{2}\right) , \ n=0, 1, 2,... \ ,  \nonumber \\
\mathcal{E}_{ym} &=& - \frac{V_{0}}{2} + \hbar\sqrt{\frac{2\gamma}{\mu_{y}}}%
\left(m + \frac{1}{2}\right) , \ m=0, 1, 2,.... \ .  \label{1daen}
\end{eqnarray}

\medskip
\par

\medskip
\par

Thus, the energy spectrum $\mathcal{E}_{nm}$ of an electron and hole
comprising a dipolar exciton in a black phosphorene double layer,
described by Eq.~(\ref{Schrel}), is
\begin{eqnarray}
\mathcal{E}_{nm} =\mathcal{E}_{xn}+\mathcal{E}_{ym}= - V_{0} + \hbar\sqrt{%
\frac{2\gamma}{\mu_{x}}}\left(n + \frac{1}{2}\right) + \hbar\sqrt{\frac{%
2\gamma}{\mu_{y}}}\left(m + \frac{1}{2}\right)  , \ n=0, 1, 2,... ;
\ m=0, 1, 2, \cdots  \ , \label{relsp}
\end{eqnarray}
while the wave function $\varphi_{nm}(x,y)$ for the relative motion
of an electron and a hole in a dipolar exciton in a black
phosphorene double layer, described by Eq.~(\ref{Schrel}), is given
by
\begin{eqnarray}
\varphi_{nm}(x,y) = \psi_{n} (x) \psi_{m} (y)  ,
\label{relwf}
\end{eqnarray}
where $\psi_{n} (x)$ and $\psi_{m} (y)$ are defined by
Eq.~(\ref{1dapsu}).
The corresponding binding energy is
\begin{eqnarray}
B = - \mathcal{E}_{00} = V_{0} - \hbar\sqrt{\frac{\gamma}{2\mu_{x}}} - \hbar%
\sqrt{\frac{\gamma}{2\mu_{y}}} = V_{0} -
\hbar\sqrt{\frac{\gamma}{2\mu_{0}}} \ . \label{bind}
\end{eqnarray}
In Eqs.~(\ref{relsp}) and~(\ref{bind})  $\mu_{0} =
\frac{\mu_{x}\mu_{y}}{\left(\sqrt{\mu_{x}}+\sqrt{\mu_{y}}
\right)^{2}}$ is ``the reduced mass of the exciton reduced masses''.
Setting  $\mu_{x} = \mu_{y} = \tilde{\mu}$  corresponding to an
isotropic system, we have $\mu_{0} = \tilde{\mu}/4$.

 We consider the phosphorene monolayers to be separated
by $h$-BN insulating layers. Besides we assume $h$-BN insulating
layers to be placed on the top and on the bottom of the phosphorene
double layer. For this insulator $\varepsilon_{d} = 4.89$ is the
effective dielectric constant, defined as $\varepsilon_{d} =
\sqrt{\varepsilon^{\bot}}\sqrt{\varepsilon^{\parallel}}$~\cite{Fogler},
where $\varepsilon^{\bot}= 6.71$ and $\varepsilon^{\parallel} =
3.56$  are the components of the dielectric tensor for
$h$-BN~\cite{CaihBN}. Since the thickness of a $h$-BN monolayer is
given by $c_{1} = 3.33 \ \mathrm{{\AA}}$~\cite{Fogler}, the
interlayer separation $D$ is presented as $D = N_{L}c_{1}$, where
$N_{L}$ is the number of $h$-BN monolayers, placed between two
phosphorene monolayers. Let us mention that $h$-BN monolayers are
characterized by relatively small density of the defects of their
crystal structure, which allowed to measure the quantum Hall effect
in the few-layer black phosphorus sandwiched between two $h$-BN
flakes~\cite{LiHBN}.

One can obtain the square of the in-plane gyration radius $r_{X}$ of
a dipolar exciton, which is the average squared projection of the
electron-hole separation onto the plane of a phosphorene
monolayer~\cite{Fogler}, as

\begin{equation}
r_{X}^{2} = \int  \varphi_{00}^{*}(x,y) (\mathbf{r}) r^{2}
\varphi_{00}(x,y) (\mathbf{r}) d^{2} r = \frac{1}{a_{x}\sqrt{\pi}}
\int_{-\infty}^{\infty} x^{2} e^{- \frac{r^{2}}{a_{x}^{2}}} d x +
\frac{1}{a_{y}\sqrt{\pi}} \int_{-\infty}^{\infty} y^{2} e^{-
\frac{y^{2}}{a_{y}^{2}}} d y   = \frac{a_{x}^{2}+ a _{y}^{2}}{2} \ .
 \label{rx2}
\end{equation}

\medskip
\par

We emphasize that the Taylor series expansion of the electron-hole
attraction potential to first order in $(r/D)^{2}$, presented in
Eq.\  (\ref{expand}) is valid if  the inequality $\left\langle r^{2}
\right\rangle = r_{X}^{2} = \left(a_{x}^{2} + a_{y}^{2}\right)/2 \ll
D^{2}$ is satisfied, where $a_{x}$ and $a_{y}$ are defined above.
Consequently, one finds that $\hbar/\left(2\sqrt{2
\mu_{0}\gamma}\right) \ll D^{2}$. The latter inequality holds for $D
\gg D_{0}$. For the Coulomb potential $D_{0} =
\hbar^{2}\varepsilon_{d}/ \left(4 ke^{2} \mu_{0}\right)$. If
$\mu_{x} = \mu_{y} = \tilde{\mu}$ for the isotropic system, we have
$D_{0} = \hbar^{2}\varepsilon_{d}/ \left(ke^{2}\tilde{\mu}\right)$.
For the Keldysh potential, one has to use Eq.\  (\ref{gamma}) for
$\gamma$ and solve the following transcendental equation

\begin{equation}
D_{0}^{3} = - \frac{\hbar^{2}\left(\varepsilon_{1}+
\varepsilon_{2}\right)\rho_{0}^{2}}{4 \pi k e^{2}\mu_{0}\left[H_{-1}\left(\frac{%
D_{0}}{\rho_{0}}\right) - Y_{-1}\left(\frac{D_{0}}{\rho_{0}}\right)
\right] }  .
 \label{KeldD0}
\end{equation}

The values of $D_{0}$ for the Keldysh and Coulomb
potentials depends on $\mu_{0}$, therefore, on the effective masses
of the electron and hole. Here and below in our calculations we use
effective masses for electron and hole from
Refs.~\onlinecite{Peng2014,Tran2014-2,Paez2014,Qiao2014}. The results, reported in these four papers,
were performed by using the first principles calculations. The different
functionals for the correlation energy and setting parameters for
the hopping lead to some difference in their results, like geometry
structures, e. g. The lattice constants in the four papers do not
coincide with each other, and this can cause the difference in the band
curvatures and effective masses. The latter motivate us to use in
calculations the different sets of masses from
Refs.~\onlinecite{Peng2014,Tran2014-2,Paez2014,Qiao2014} that allows
to understand the dependence of the binding energy, the sound
velocity, and the mean field temperature of the superfluidity on effective masses of electrons and holes.

\begin{figure}[h]
\centering
\includegraphics[width=18.0cm]{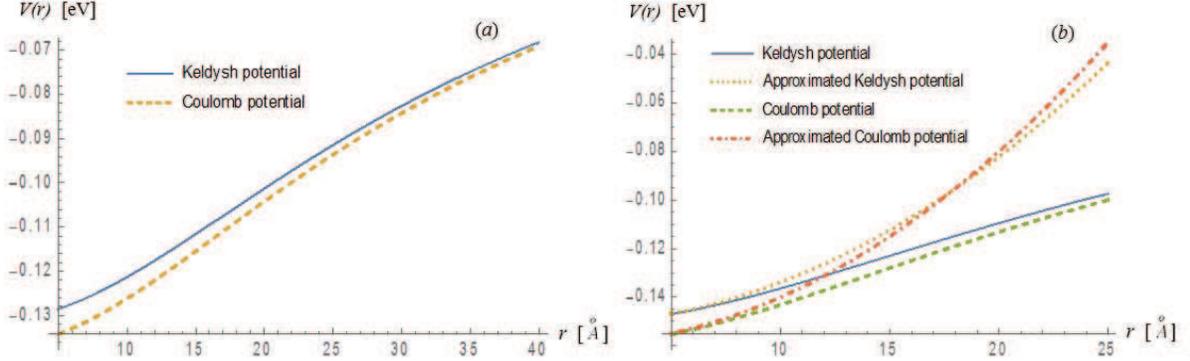} \vspace{-2.7cm}
\caption{(Color online)   (a) The Keldysh and Coulomb potentials for
electron-hole attraction in a black  phosphorene double layer. (b)
Comparison of the Keldysh and Coulomb electron-hole attractions in a
black phosphorene double layer approximated by the harmonic
oscillator potential. The calculations were performed for the number
$N_{L} = 7$ of $h$-BN monolayers, placed between two phosphorene
monolayers, the set of masses from Ref. \protect\cite{Peng2014} and
polarizability from Ref. \protect\cite{Rodin}.} \label{Fig2}
\end{figure}

The values of $D_{0}$ for  the Keldysh potential,
obtained by solving Eq.\ (\ref{KeldD0}), and the Coulomb potential
for the sets of the masses from
Refs.~[\onlinecite{Peng2014,Tran2014-2,Paez2014,Qiao2014}],
respectively, are given in Table~\ref{tab1}.  As it can be seen in
Table~\ref{tab1}, the characteristic value of
 $D_{0}$, entering the condition $D \gg D_{0}$ of validity
of the first order Taylor expansion of electron-hole attraction
potential, given by Eq.\  (\ref{expand}), is about one order of
magnitude smaller for the Keldysh potential than for the Coulomb
potential. Therefore, the first order Taylor   expansion can be
valid for the smaller interlayer separations $D$  for the Keldysh
potential than for the Coulomb potential. Thus, validity of the
harmonic oscillator approximation of the Keldysh potential is more
reasonable. This is due to the fact that the Keldysh potential
describes the screening, which makes the Keldysh potential to be
more short-range than the Coulomb potential. Therefore, the harmonic
oscillator approximation of electron-hole attraction potential,
given by Eq.\ (\ref{expand}), can be valid for smaller number
$N_{L}$ of $h$-BN insulating layers between two phosphorene
monolayers for the Keldysh potential than for the Coulomb potential.
According to Table~\ref{tab1}, for both potentials $D_{0}$ is not
sensitive to the choice of the set of effective electrons and holes
masses. Comparisons of the Keldysh and Coulomb interaction
potentials for an electron-hole pair and their approximations using
harmonic oscillator potentials obtained from a Taylor series
expansion are presented in Fig. \ref{Fig2}. According to Fig.
\ref{Fig2}a, the Keldysh potential is weaker than the Coulomb
potential at small  projections $r$ of the electron-hole distance on
the phosphorene monolayer plane, while the both potentials become
closer to each other as $r$ increases, demonstrating almost no
difference at $r \gtrsim 25 \ {\AA}$.

\medskip
\par
\begin{table}[t]
\caption{Value for $D_{0}$ for Keldysh and Coulomb potentials for different sets of masses for electron and hole from Refs. \cite{Peng2014}, \cite%
{Tran2014-2}, \cite{Paez2014}, and \cite{Qiao2014}. }
\begin{center}
 \begin{tabular}{cccccc}
\hline\hline Mass from Ref: &  & \cite{Peng2014} & \cite{Tran2014-2}
& \cite{Paez2014} & \cite{Qiao2014} \\ \hline
Keldysh potential & $D_{0},\ \mathrm{{\AA}}$ & 1.0 & 0.98 & 0.9 & 0.9 \\
Coulomb potential & $D_{0},\ \mathrm{{\AA}}$ & 14.7 & 14.4 & 12.2 & 12.3 \\
\hline\hline
\end{tabular}
\end{center}
\label{tab1}
\end{table}

For the number $N_{L} = 7$ of $h$-BN monolayers, placed between two
phosphorene monolayers, the binding energies of dipolar excitons,
 calculated for  the sets of the masses from
Refs.~[\onlinecite{Peng2014,Tran2014-2,Paez2014,Qiao2014}] by using
Eq.~(\ref{bind}), are given by   $28.2 \ \mathrm{meV}$, $ 29.6 \
\mathrm{meV}$, $ 37.6 \ \mathrm{meV}$, and $ 37.2 \ \mathrm{meV}$.
Let us mention that the maximal dipolar exciton binding energy was
obtained for the set of the masses, taken from
Ref.~[\onlinecite{Qiao2014}].  The  dipolar exciton binding energy
increases when  the reduced mass $\mu_{0}$ of the exciton reduced
masses increases. The reduced mass $\mu_{0}$
 for the sets of the masses from
Refs.~[\onlinecite{Peng2014,Tran2014-2,Paez2014,Qiao2014}] is
presented in Table~\ref{tab2}. One can conclude that while $D_{0}$
is not sensitive to the choice of the set of effective electrons and
holes masses, the binding energy of indirect exciton
depends on the exciton reduced mass $\mu_{0}$, which is
defined by the effective electron and hole masses.

\medskip
\par

It is worthy of note that the energy spectrum of the center-of-mass
of an electron-hole pair $\varepsilon_{0}(\mathbf{P})$   may be
expressed as

\begin{eqnarray}
\varepsilon_{0}(\mathbf{P}) = \frac{P_{x}^{2}}{2M_{x}} + \frac{P_{y}^{2}}{%
2M_{y}}  .
\label{eps0}
\end{eqnarray}
Substituting the polar coordinate for the momentum $P_{x} = P \cos
\Theta$ and $P_{y} = P \sin \Theta$ into Eq.~(\ref{eps0}), we obtain

\begin{eqnarray}
\varepsilon_{0}(\mathbf{P}) = \varepsilon_{0}(P,\Theta) = \frac{P^{2}}{%
2M_{0}(\Theta)},
 \label{eps0pol}
\end{eqnarray}
where $M_{0}(\Theta)$ is the effective angle-dependent exciton mass in a
black phosphorene double layer, given by

\begin{eqnarray}
M_{0}(\Theta) = \left[\frac{\cos^{2}\Theta}{M_{x}} + \frac{\sin^{2}\Theta}{%
M_{y}} \right]^{-1}.
 \label{M0}
\end{eqnarray}

\section{Collective excitations for dipolar excitons in a black phosphorene
double layer}
\label{collect}

We now turn our attention  to a dilute distribution of electrons and
holes in a pair of parallel black phosphorene layers spatially
separated by a dielectric, when $n r_{X}^{2}\ll 1$, where $n$  is
the concentration  for dipolar excitons. In this limit, the dipolar
excitons are formed by electron-hole pairs with the electrons and
holes spatially separated in two different phosphorene layers.

\medskip
\par

The distinction between excitons, which are not an elementary but a
composite bosons~\cite{Comberscot} and bosons is caused by exchange
effects~\cite{Moskalenko_Snoke}. At large interlayer separations
$D$,  the exchange effects in the exciton-exciton interactions in a
phosphorene double layer can be neglected, since the exchange
interactions in a spatially separated electron-hole system in a
double layer are suppressed due to the low tunneling probability,
caused by the shielding of the dipole-dipole interaction by the
insulating barrier~\cite{BK,BLG}. Therefore, we treat the dilute
system of dipolar excitons in a phosphorene double layer as a weakly
interacting Bose gas.

\medskip
\par

The model Hamiltonian $\hat{H}$ of the 2D interacting dipolar excitons is given by

\begin{eqnarray}
\hat{H}= \sum_{\mathbf{P}} \varepsilon_{0}(P,\Theta)a_{\mathbf{P}}^{\dagger
}a_{\mathbf{P}} + \frac{g}{S}\sum_{\mathbf{P}_{1}\mathbf{P}_{2}\mathbf{P}%
_{3}}a_{\mathbf{P}_{1}}^{\dagger}a_{\mathbf{P}_{2}}^{\dagger}a_{\mathbf{P}%
_{3}}a_{\mathbf{P}_{1}+\mathbf{P}_{2}-\mathbf{P}_{3}} ,
\label{Ham}
\end{eqnarray}
where $a_{\mathbf{P}}^{\dagger }$ and $a_{\mathbf{P}}$ are Bose creation and
annihilation operators for dipolar excitons with momentum $\mathbf{P}$,
$S$ is a  normalization area for the system, $\varepsilon_{0}(P,\Theta)$ is the
angular-dependent energy spectrum of non-interacting dipolar excitons, given
by Eq.\  (\ref{eps0pol}), and $g$ is a coupling constant for the interaction
between two dipolar excitons.

\medskip
\par

We expect that at $T=0$ K almost all dipolar excitons condense into
a BEC.  One can treat this weakly interacting gas of
dipolar excitons within the Bogoliubov
approximation~\cite{Abrikosov,Lifshitz}. The Bogoliubov
approximation for a weakly interacting Bose gas allows us to
diagonalize the many-particle Hamiltonian, replacing the product of
four operators in the interaction term by the product of two
operators. This is justified under the assumption that most of the
particles belong to the BEC, and only the interactions between the
condensate and non-condensate particles are taken into account,
while the interactions between non-condensate particles are
neglected. The condensate operators are replaced by
numbers \cite{Abrikosov}, and the resulting Hamiltonian is quadratic
with respect to the creation and annihilation operators. Employing
the Bogoliubov approximation~\cite{Lifshitz}, we obtain the chemical
potential $\mu$ of the entire exciton system by minimizing
$\hat{H}_{0}-\mu \hat{N}$ with respect to the 2D concentration $n$,
where $\hat{N}$ denotes the number operator.  The later one is

\begin{eqnarray}
\hat{N}=\sum_{\mathbf{k}}a_{\mathbf{P}}^{\dagger }a_{\mathbf{P}},
\label{Nop}
\end{eqnarray}
while $H_{0}$ is the Hamiltonian describing the particles in the
condensate with zero momentum $\mathbf{P}=0$.  The minimization of
$\hat{H}_{0}-\mu \hat{N}$ with respect to the number of excitons
$N=Sn$ results in the standard expression~\cite{Abrikosov,Lifshitz}

\begin{eqnarray}
\mu = g n .  \label{mu1}
\end{eqnarray}

Following the procedure presented in Ref.\ [\onlinecite{BKKL}], the
interaction parameters for the exciton-exciton interaction in very
dilute systems could be obtained assuming the exciton-exciton
dipole-dipole repulsion exists only at distances between excitons
greater than distance from the exciton to the classical turning
point. The distance between two excitons cannot be less than this
distance, which is determined by the conditions reflecting the fact
that the energy of two excitons cannot exceed the doubled chemical
potential $\mu $ of the system, i.e.,

\begin{eqnarray}
U(R_{0})= 2\mu . \label{cond}
\end{eqnarray}
In Eq.~(\ref{cond})   $U(R_{0})$ is the potential of interaction
between two dipolar excitons at the distance $R_{0}$, where $R_{0}$
corresponds to the distance between two dipolar excitons at their
classical turning point.

For our model we investigate the formation of dipolar excitons in a
phosphorene double layer with the use of the Keldysh and Coulomb interactions.
Therefore, it is reasonable to adopt the general approach for
treating collective excitations of dipolar excitons. If the distance
between two dipolar excitons is $R$ and the electron and hole of one
dipolar exciton interact with the electron and hole of the other
dipolar exciton, it is straightforward to show that the
exciton-exciton interaction $U(R)$ has the form:

\begin{equation}
U(R)=2V(R)-2V\left(R\sqrt{1+\frac{D^{2}}{R^{2}}}\right),
\label{Keldysh Dipole}
\end{equation}%
where $V(R)$ represents the interaction potential between two
electrons or two holes in the same phosphorene monolayer. We can
assume the potential $V(R)$ to be given by either Keldysh potential
(\ref{Keldysh}) or by Coulomb potential.

In a very dilute system of dipolar excitons and, therefore, $D \ll
R$, one may expand the second term in Eq. (\ref{Keldysh Dipole}) in
terms of $(D/R)^{2},$ and by retaining only the first order terms
with respect to $(D/R)^{2}$, finally obtains

\begin{eqnarray}   \label{Dipolar Keldysh Approx}
U(R)=\left\{
\begin{array}{c}
\frac{\pi ke^{2}D^{2}}{2\varepsilon _{d}\rho _{0}^{2}R}\left[
Y_{-1}\left(
\frac{R}{\rho _{0}}\right) -H_{-1}(y)\left( \frac{R}{\rho _{0}}\right) %
\right],\text{ for the Keldysh potential, } \\
\frac{ke^{2}D^{2}}{\epsilon _{d}R^{3}},\text{ \ \ \ \ \ \ \ \ \ \ \
\ \ \ \ \ \ \ \ \ \ \ \ \ \ \ \ \ \ \ \ \ \ \ \ \ \ \   for the
Coulomb potential.}%
\end{array}%
\right.
\end{eqnarray}

\medskip
\par

Following the procedure presented in Ref.\ [\onlinecite{BKKL}], one
can obtain the coupling constant for the exciton-exciton
interaction:
\begin{eqnarray}
g =   2\pi \int_{R_{0}}^{\infty} R dR\ U(R)
  . \label{gK1}
\end{eqnarray}

Substituting Eq.~~(\ref{Dipolar Keldysh Approx}) into
Eq.~~(\ref{gK1}), one obtains the exciton-exciton coupling
constant $g$  as following
\begin{eqnarray}  \label{gK}
g=\left\{
\begin{array}{c}
\frac{2\pi ^{2}ke^{2}D^{2}}{2\epsilon _{d}\rho _{0}}\left[ H_{0}\left( \frac{%
R_{0}}{\rho _{0}}\right) -Y_{0}\left( \frac{R_{0}}{\rho _{0}}\right)
\right]
,\text{ for the Keldysh potential,} \\
\frac{2\pi ke^{2}D^{2}}{\epsilon _{d}R_{0}},\text{ \ \ \ \ \ \ \ \ \
\ \ \ \
\ \ \ \ \ \ \ \ \ \ \ \ \ \ \ \ \ \ \   for the Coulomb potential.%
}%
\end{array}%
\right.
\end{eqnarray}

Combining Eqs.\ (\ref{cond}),~(\ref{Dipolar Keldysh Approx})
and~(\ref{gK}), for the Keldysh potential we obtain the following
equation for $R_{0}$:
\begin{eqnarray}
4 \pi n \rho_{0}^{2} y \left[ H_{0}(y)-Y_{0}(y)\right] = - \left[
H_{-1}(y)-Y_{-1}(y)\right]
  , \label{R0K}
\end{eqnarray}
where $y = R_{0}/\rho_{0}$.

Combining Eqs.\ (\ref{cond}),~(\ref{Dipolar Keldysh Approx})
and~(\ref{gK}),  we obtain the following expression for $R_{0}$ in the case of Coulomb
potential

\begin{eqnarray}
R_{0} =  \frac{1}{2\sqrt{\pi n}}  . \label{r0}
\end{eqnarray}

 From Eqs.\ (\ref{r0}),~(\ref{gK}) and~(\ref{mu1}), one obtains the
exciton-exciton coupling constant $g$ for the Coulomb potential

\begin{eqnarray}
g=\frac{4\pi ke^{2}D^{2}\sqrt{\pi n}}{\epsilon _{d}} .
\label{geqeq1}
\end{eqnarray}

\medskip
\par

\begin{figure}[h]
\centering
\includegraphics[width=15.0cm]{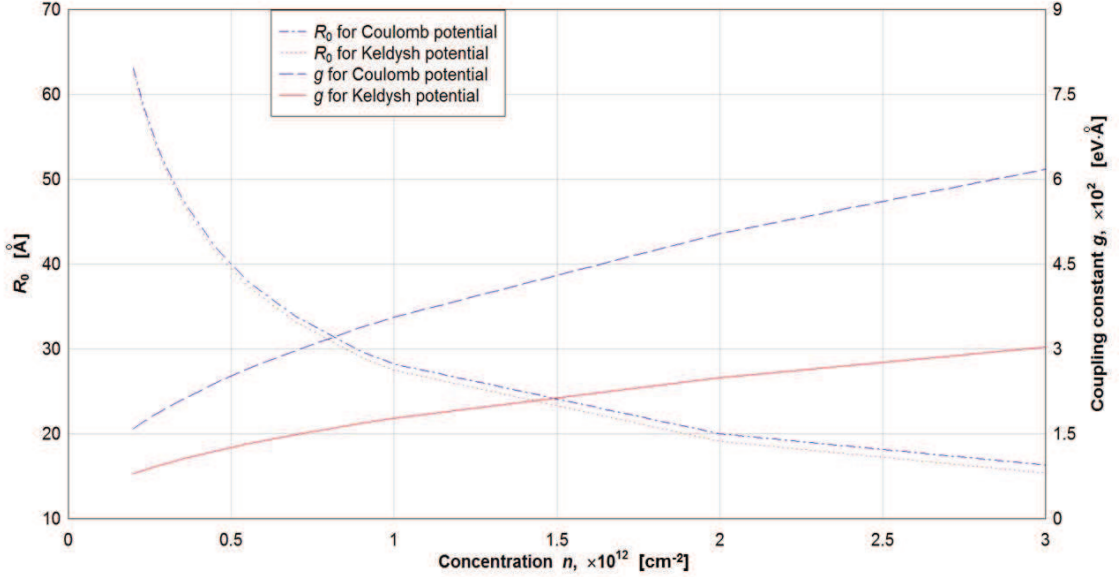}
\caption{(Color online) The coupling constant $g$ and  the
distance $R_{0}$ between two dipolar excitons at the classical
turning point for the Keldysh and Coulomb potentials for a
phosphorene double layer as functions of the exciton concentration.
The number of $h$-BN monolayers between the phosphorene monolayers
is $N_{L} = 7$.} \label{Figg}
\end{figure}

The coupling constant $g$ and  the distance $R_{0}$ between two
dipolar excitons at the classical turning point for the Keldysh and
Coulomb potentials for a phosphorene double layer as functions of
the exciton concentration are represented in Fig.~\ref{Figg}.
According  to Fig.~\ref{Figg}, $R_{0}$ decreases with the increase
of the exciton concentration $n$. While for the Coulomb potential
$R_{0}$ is slightly larger than for the Keldysh potential, the
difference is very small. As shown in Fig.~\ref{Figg}, the
coupling constant $g$  is larger for the Coulomb potential than
for  the Keldysh potential, because the interaction between the
charge carriers, interacting via the Kledysh potential, is
suppressed by the screening effects. The difference between $g$  for
the Keldysh and Coulomb potentials increases as the exciton
concentration $n$ increases.

\medskip
\par

The many-particle Hamiltonian of dipolar excitons in a black
phosphorene double layer given by Eq.\  (\ref{Ham}) is standard
for a weakly interacting Bose gas with the only  difference being that the
single-particle energy spectrum of non-interacting excitons is
angular-dependent due to the orientation variation of the exciton effective
mass. Whereas the first term in Eq.\  (\ref{Ham}) which is responsible for the
single-particle kinetic energy is angular dependent, the second interaction
term in Eq.\  (\ref{Ham}) does not depend on an angle because the
dipole-dipole repulsion between excitons does not depend on an angle.
Therefore, for a weakly interacting gas of dipolar excitons in a black
phosphorene double layer, in the framework of the Bogoliubov approximation, we could apply exactly the same procedure which has been adapted for a standard weakly
interacting Bose gas \cite{Abrikosov,Lifshitz}, but taking into account the
angular dependence of a single-particle energy spectrum of dipolar excitons.
Therefore, the Hamiltonian $\hat{H}_{col}$ of the collective excitations in
the Bogoliubov approximation for the weakly interacting gas of dipolar
excitons in black phosphorene is given by

\begin{eqnarray}
\hat{H}_{col} = \sum_{P \neq 0,\Theta}\varepsilon(P,\Theta)\alpha_{\mathbf{P}%
}^{\dagger}\alpha_{\mathbf{P}} ,
\label{Hamq}
\end{eqnarray}
where $\alpha_{j\mathbf{P}}^{\dagger}$ and $\alpha_{j\mathbf{P}}$ are the
creation and annihilation Bose operators for the quasiparticles with the
energy dispersion corresponding to the angular dependent spectrum of the
collective excitations $\varepsilon(P,\Theta)$, described by

\begin{eqnarray}
\varepsilon(P,\Theta) = \left[ \left(\varepsilon_{0}(P,\Theta) +
gn\right)^{2} - \left(gn\right)^{2}\right]^{1/2}  .
\label{colsp}
\end{eqnarray}

\medskip
\par

In the limit of small momenta $P$, when $\varepsilon_{0}(P,\Theta)
\ll gn$, we expand the spectrum of collective excitations
$\varepsilon(P,\Theta)$ up to first order with respect to the
momentum $P$ and obtain the sound mode of the collective excitations
$\varepsilon(P,\Theta) = c_{S}(\Theta) P $, where $c_{S}(\Theta)$ is
the angular dependent sound velocity, given by

\begin{eqnarray}
c_{S}(\Theta)=\sqrt{\frac{gn}{M_{0}(\Theta)} } .  \label{cs}
\end{eqnarray}

\medskip
\par

\begin{figure}[h]
\centering
\includegraphics[width=18.0cm]{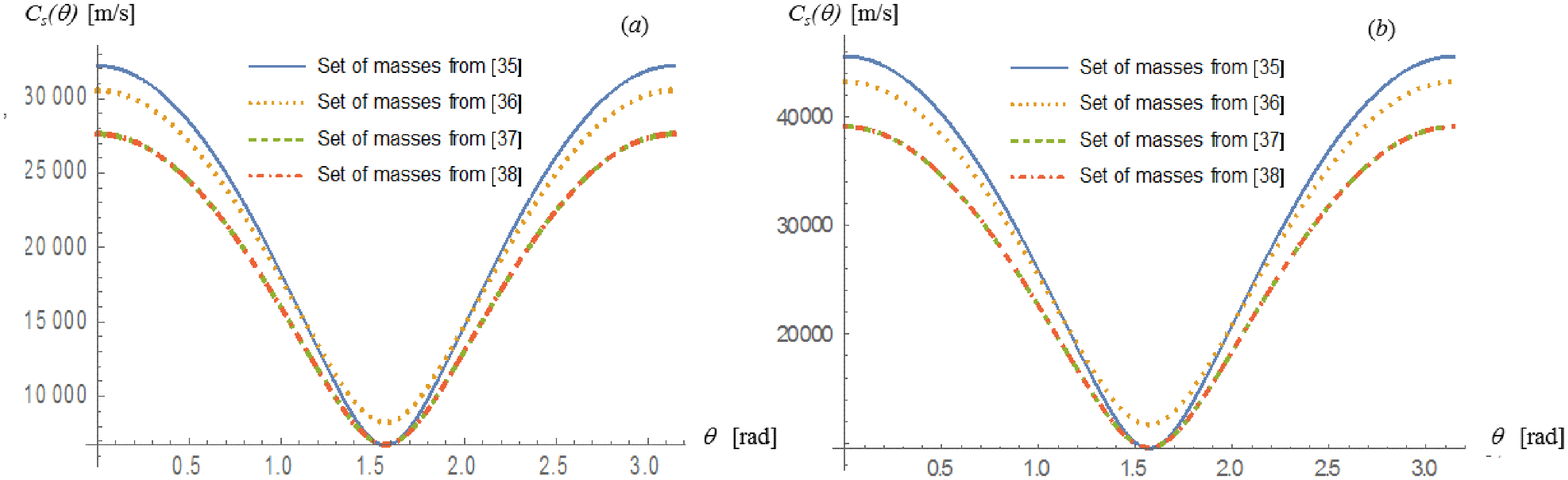}  \vspace{-2.7cm}
\caption{(Color online) The angular dependence of the sound
velocity. (a) The interaction between the carriers is described by
the Keldysh potential. (b) The interaction between the carriers is
described by the Coulomb potential. The calculations were performed
for the exciton concentration $n= 2 \times 10^{16} \
\mathrm{m^{-2}}$ and the number $N_{L} = 7$ of $h$-BN monolayers,
placed between two phosphorene monolayers.} \label{Fig4}
\end{figure}

The asymmetry of the electron and hole dispersion in black
phosphorene is reflected in the angular dependence of the sound
velocity through the angular dependence of the effective exciton
mass.  The angular dependence of the sound velocity for the Keldysh
and Coulomb potentials  is presented in Fig.~\ref{Fig4}, where it is
demonstrated that the exciton sound velocity  is maximal at $\Theta
= 0$ and $\Theta = \pi$ and minimal at $\Theta = \pi/2$. As it
follows from comparison of Fig.~\ref{Fig4}a with Fig.~\ref{Fig4}b,
at the same parameters, the sound velocity $c_{S}(\Theta)$ is
greater in the case of Coulomb potential for the
interaction between the charge carriers than for the Keldysh
potential, because the Keldysh potential implies the screening
effects, which make the interaction between the carriers weaker.
According to Fig.~\ref{Fig4}, the sound velocity depends on the
effective electron and hole masses. However, the sound velocities
are coincided at all angles $\Theta$ for two sets of masses from
Refs.~[\onlinecite{Paez2014}] and~[\onlinecite{Qiao2014}],
correspondingly. Since at low momenta the sound-like energy spectrum
of collective excitations in the dipolar exciton system in a
phosphorene double layer satisfies  to the Landau
criterion for superfluidity, the dipolar exciton superfluidity in a
black phosphorene double layer is possible.  Let us mention that the
exciton concentration, used for the calculations, represented in
Fig.~\ref{Fig4} and below,  corresponds by the order of
magnitude to the experimental values~\cite{Warren,Surrente}.

\medskip
\par

\section{Superfluidity of dipolar excitons in a black phosphorene double Layer}
\label{super}

\medskip
\par

Since, at small momenta, the energy spectrum of the quasiparticles
for  a weakly interacting gas of dipolar excitons is sound-like,
this means that the system satisfies to the Landau
criterion for superfluidity \cite{Abrikosov,Lifshitz}. The critical
exciton velocity for superfluidity is angular-dependent, and it is
given by $v_{c}(\Theta)= c_{S}(\Theta)$, because the quasiparticles
are created at velocities above the angle dependent velocity of
sound.
 According to Fig.~\ref{Fig4}, the critical exciton
velocity for superfluidity has maximum at $\Theta = 0$ and $\Theta =
\pi$ and has minimum at $\Theta = \pi/2$. Therefore, as shown in
Fig.~\ref{Fig4}a, if the excitons move with the velocities in the
range of approximately between $8 \times 10^{3} \ \mathrm{m/s}$ and
$3.4 \times 10^{4} \ \mathrm{m/s}$, the superfluity is present for
the angles at the edges of the angle range between $\Theta = 0$ and
$\Theta = \pi$, while the superfluidity  is absent at the center of
this angle range.

\medskip
\par

The density of the superfluid component $\rho_{s}(T)$ is defined as
$\rho_{s}(T) = \rho - \rho_{n}(T)$, where $\rho$ is the total 2D density of
the system and $\rho_{n}(T)$ is the density of the normal component. We
define the normal component density $\rho_{n}(T)$ in the usual
way.\cite{Pitaevskii}. Suppose that the excitonic system moves with a velocity
$\mathbf{u}$, which means that the superfluid component moves with the
velocity $\mathbf{u}$. At nonzero temperatures $T$ dissipating
quasiparticles will appear in this system. Since their density is small at
low temperatures, one may assume that the gas of quasiparticles is an ideal
Bose gas. To calculate the superfluid component density, we define the total
mass current $\mathbf{J}$ for a Bose-gas of quasiparticles in the frame of reference
where the superfluid component is at rest, by

\begin{eqnarray}
\mathbf{J}=\int \frac{s d^{2}P}{(2\pi \hbar )^{2}}\mathbf{P} f\left[
\varepsilon(P,\Theta)-\mathbf{P}\mathbf{u}\right] . \label{nnor}
\end{eqnarray}
In Eq.\ (\ref{nnor})  $f\left[ \varepsilon(P,\Theta)\right] =\left( \exp \left[
\varepsilon(P,\Theta)/(k_{B}T)\right] -1\right)^{-1}$ is the
Bose-Einstein distribution function for quasiparticles with the
angule dependent dispersion $\varepsilon(P,\Theta)$, $s = 4$
is the spin degeneracy factor, and  $k_{B}$ is the Boltzmann
constant. Expanding the integrand of   Eq.\ (\ref{nnor}) in terms of
$\mathbf{P}\mathbf{u}/(k_{B}T)$ and restricting ourselves by
 the first order term, we obtain

\begin{eqnarray}
\mathbf{J}= - \frac{s}{k_{B} T}\int \frac{d^{2}P}{(2\pi \hbar )^{2}}
\mathbf{P}\left(\mathbf{Pu}\right) \frac{\partial f\left[
\varepsilon(P,\Theta)\right] }{\partial \varepsilon(P,\Theta)}  .
\label{J_Tot}
\end{eqnarray}
The normal density $\rho_{n}$ in the anisotropic system has tensor form \cite{Saslow}.
We define the tensor elements for the normal component density
$\rho_{n}^{(ij)}(T)$ by

\begin{eqnarray}
J_{i} = \rho_{n}^{(ij)}(T) u _{j}  ,
\label{rhodef}
\end{eqnarray}
where $i$ and $j$ denote either the  $x$ or $y$ component of the
vectors. Assuming that  the vector ${\bf u}\uparrow\uparrow OX$
($\uparrow\uparrow$ denotes that ${\bf u}$ is parallel to the $OX$
axis and has the same direction as the $OX$ axis),
 we have ${\bf u} = u_{x}{\bf i}$
and ${\bf P} = P_{x}{\bf i} + P_{y}{\bf j}$. Therefore, we obtain

\begin{eqnarray}
{\bf P}\cdot {\bf u}  &=& P_{x}u_{x},
\nonumber \\
{\bf P}\left({\bf P}\cdot{\bf u}\right) &=& P_{x}^{2}u_{x}{\bf i} +
P_{x}P_{y}u_{x}{\bf j} ,
\label{i2}
\end{eqnarray}
where ${\bf i}$ and ${\bf j}$ are unit vectors in the $x$ and $y$
directions, respectively. Upon substituting Eq.\ (\ref{i2}) into
Eq.\ (\ref{J_Tot}), one obtains

\begin{eqnarray}
J_{x}=  - \frac{s}{k_{B} T}\int_{0}^{\infty}dP
\frac{P^{3}}{(2\pi \hbar )^{2}} \int_{0}^{2\pi} d \Theta \frac{\partial f%
\left[ \varepsilon(P,\Theta)\right] }{\partial \varepsilon(P,\Theta)}
\cos^{2}\Theta u_{x}  .
\label{Jx}
\end{eqnarray}
Using the definition of the density for the normal component from
Eq.\ (\ref{rhodef}), we obtain

\begin{eqnarray}
\rho_{n}^{(xx)}(T) = \frac{s}{k_{B} T}\int_{0}^{\infty} dP\frac{
P^{3}  }{(2\pi \hbar )^{2}}
\int_{0}^{2\pi} d \Theta \frac{\exp \left[ \varepsilon(P,\Theta)/(k_{B}T)%
\right]}{\left( \exp \left[ \varepsilon(P,\Theta)/(k_{B}T)\right]
-1\right)^{2}}\cos^{2}\Theta  .
\label{rhonx}
\end{eqnarray}
Substitution of Eq.\ (\ref{i2}) into Eq.\ (\ref{J_Tot}) gives

\begin{eqnarray}
J_{y}&=& - \frac{s}{k_{B} T}\int \frac{d^{2}P}{(2\pi \hbar )^{2}}
P_{x}P_{y} \frac{\partial f\left[ \varepsilon(P,\Theta)\right]
}{\partial \varepsilon(P,\Theta)} u_{x}
\nonumber \\
&=& \frac{s}{k_{B} T}\int_{0}^{\infty}dP \frac{P^{3}}{(2\pi \hbar
)^{2}}
\int_{0}^{2\pi} d \Theta \frac{\exp \left[ \varepsilon(P,\Theta)/(k_{B}T)%
\right]}{\left( \exp \left[ \varepsilon(P,\Theta)/(k_{B}T)\right]
-1\right)^{2}} \cos\Theta \sin\Theta u_{x}= 0  .
\label{Jy}
\end{eqnarray}
The integral in Eq.~(\ref{Jy}) equals to zero,  since the integral
over the angle $\Theta$
over the period of the function results in zero. Therefore, one obtains $%
\rho_{n}^{(xy)} = 0$.

\medskip
\par

Now assuming the vector ${\bf u}\uparrow\uparrow OY$,
 we obtain analogously the following relations:

\begin{eqnarray}
\rho_{n}^{(yy)}(T) &=& \frac{s}{k_{B} T}\int_{0}^{\infty} dP \frac{ P^{3}}{%
(2\pi \hbar )^{2}} \int_{0}^{2\pi} d \Theta \frac{\exp \left[
\varepsilon(P,\Theta)/(k_{B}T)\right]}{\left( \exp \left[ \varepsilon(P,%
\Theta)/(k_{B}T)\right] -1\right)^{2}} \sin^{2} \Theta  ,  \nonumber \\
\rho_{n}^{(yx)}(T) &=& 0 \ .
 \label{rhony}
\end{eqnarray}

\medskip
\par

By defining  the tensor of the concentration of the normal component as the
linear response of the flow of quasiparticles on the external velocity as $%
n_{n}^{(ij)} = \rho_{n}^{(ij)}/M_{i}$, one obtains:

\begin{eqnarray}
n_{n}^{(xx)}(T) &=& \frac{s}{k_{B} M_{x}T}\int_{0}^{\infty}dP \frac{
P^{3} }{(2\pi \hbar )^{2}} \int_{0}^{2\pi} d \Theta \frac{\exp
\left[
\varepsilon(P,\Theta)/(k_{B}T)\right]}{\left( \exp \left[ \varepsilon(P,%
\Theta)/(k_{B}T)\right] -1\right)^{2}} \cos^{2} \Theta ,  \nonumber \\
n_{n}^{(xy)}(T) &=& 0 \   \nonumber \\
n_{n}^{(yy)}(T) &=& \frac{s}{k_{B} M_{y}T}\int_{0}^{\infty}dP \frac{
P^{3} }{(2\pi \hbar )^{2}} \int_{0}^{2\pi} d \Theta \frac{\exp
\left[
\varepsilon(P,\Theta)/(k_{B}T)\right]}{\left( \exp \left[ \varepsilon(P,%
\Theta)/(k_{B}T)\right] -1\right)^{2}} \sin^{2} \Theta  ,  \nonumber \\
n_{n}^{(yx)}(T) &=& 0 .  \label{nnxy}
\end{eqnarray}

The linear response of the flow of quasiparticles $\mathbf{J}_{qp}$
with respect to the external velocity at any angle measured from the
$OX$ direction is given in terms of the angle dependent
concentration for the normal component $\tilde{n}_{n}(\Theta, T)$
 as

\begin{eqnarray}
\left|\mathbf{J}_{qp}\right| &=& \left|n_{n}^{(xx)}(T)u_{x}{\bf i} +
n_{n}^{(yy)}(T)u_{y}{\bf j}\right|  \nonumber \\
&=& \sqrt{\left[n_{n}^{(xx)}(T)\right]^{2}u^{2}\cos^{2} \Theta + \left[%
n_{n}^{(yy)}(T)\right]^{2} u^{2} \sin^{2} \Theta } =
\tilde{n}(\Theta, T)u , \label{nabs}
\end{eqnarray}
where the concentration of the normal component
$\tilde{n}_{n}(\Theta, T)$ is

\begin{eqnarray}
\tilde{n}_{n}(\Theta, T) =
\sqrt{\left[n_{n}^{(xx)}(T)\right]^{2}\cos^{2} \Theta +
\left[n_{n}^{(yy)}(T)\right]^{2} \sin^{2} \Theta }  . \label{nnang}
\end{eqnarray}
From Eq.~(\ref{nnang}) it follows that $n_{n}^{(xx)} = \tilde{n}_{n}(\Theta =
0)$ and $n_{n}^{(yy)} = \tilde{n}_{n}(\Theta = \frac{\pi}{2})$.

Eq.~(\ref{nnang})  can be rewritten in the following form:
\begin{eqnarray}
\tilde{n}_{n}(\Theta, T) =
\sqrt{\frac{\left[n_{n}^{(xx)}(T)\right]^{2} +
\left[n_{n}^{(yy)}(T)\right]^{2}}{2} +
\frac{\left(\left[n_{n}^{(xx)}(T)\right]^{2} -
\left[n_{n}^{(yy)}(T)\right]^{2}\right)\cos\left( 2 \Theta
\right)}{2}}  . \label{nnang1}
\end{eqnarray}

\medskip
\par

We define the angle dependent concentration of the superfluid
component $\tilde{n}_{s}(\Theta, T)$ by
\begin{eqnarray}
\tilde{n}_{s}(\Theta, T) = n - \tilde{n}_{n}(\Theta, T)  ,
\label{superfli}
\end{eqnarray}
where $n$ is the total concentration of the dipolar excitons. The
mean field critical temperature $T_{c}(\Theta)$ of the phase
transition related to the occurrence of superfluidity in the
direction with the angle $\Theta$ relative to the $x$ direction is
determined by the condition

\begin{eqnarray}
\tilde{n}_{n}(\Theta, T_{c}(\Theta)) = n  . \label{Tc}
\end{eqnarray}

\subsection{Superfluidity for the sound-like spectrum
of collective excitations}

\medskip
\par

For small momenta,  substituting  the sound spectrum of collective
excitations $\varepsilon(P,\Theta) = c_{S}(\Theta) P $ with the
angular-dependent sound velocity $c_{S}(\Theta)$, given by
Eq.~(\ref{cs}), into Eq.~(\ref{nnxy}), we obtain
\begin{eqnarray}
n_{n}^{(xx)}(T) &=& \frac{2s(k_{B}T)^{3}\zeta (3)}{(\pi
\hbar)^{2}M_{x}}  \int_{0}^{2\pi} \frac{\cos^{2}
\Theta}{c_{S}^{4}(\Theta)} d \Theta =
 \frac{2s(k_{B}T)^{3}\zeta (3)}{(\pi \hbar
g n)^{2}M_{x}}  \int_{0}^{2\pi} \frac{\cos^{2} \Theta}{\left(\frac{\cos^{2}\Theta}{M_{x}} + \frac{\sin^{2}\Theta}{%
M_{y}} \right)^{2} } d \Theta
 ,  \nonumber \\
n_{n}^{(xy)}(T) &=& 0 ,   \nonumber \\
n_{n}^{(yy)}(T) &=& \frac{2s(k_{B}T)^{3}\zeta (3)}{(\pi \hbar
)^{2}M_{y}} \int_{0}^{2\pi} \frac{\sin^{2}
\Theta}{c_{S}^{4}(\Theta)} d \Theta =  \frac{2s(k_{B}T)^{3}\zeta
(3)}{(\pi \hbar g n)^{2}M_{y}}  \int_{0}^{2\pi} \frac{\sin^{2}
\Theta}{\left(\frac{\cos^{2}\Theta}{M_{x}} + \frac{\sin^{2}\Theta}{%
M_{y}} \right)^{2} } d \Theta  ,  \nonumber \\
n_{n}^{(yx)}(T) &=& 0  ,  \label{nnxy00}
\end{eqnarray}
where  $\zeta (z)$ is the Riemann zeta function ($\zeta (3)\simeq
1.202$).

The integrals in  Eq.~(\ref{nnxy00}) can be evaluated analytically.
Substituting the following expressions
\begin{eqnarray}
\int_{0}^{2\pi} \frac{\cos^{2} \Theta}{\left(\frac{\cos^{2}\Theta}{M_{x}} + \frac{\sin^{2}\Theta}{%
M_{y}} \right)^{2} } d \Theta  = \pi M_{x}\sqrt{M_{x}M_{y}}  ,
\hspace{2cm} \int_{0}^{2\pi} \frac{\sin^{2}
\Theta}{\left(\frac{\cos^{2}\Theta}{M_{x}} + \frac{\sin^{2}\Theta}{%
M_{y}} \right)^{2} } d \Theta  = \pi M_{y}\sqrt{M_{x}M_{y}}  ,
\label{angint}
\end{eqnarray}
into Eq.~(\ref{nnxy00}), one obtains
\begin{eqnarray}
n_{n}^{(xx)}(T) = n_{n}^{(yy)}(T) = \frac{2  \zeta
(3)s(k_{B}T)^{3}\sqrt{M_{x}M_{y}}}{\pi (\hbar g n)^{2}}  ,
\hspace{2cm} n_{n}^{(xy)}(T) = n_{n}^{(yx)}(T)  = 0  .
\label{nnxyfinal}
\end{eqnarray}
Let us mention that Eq.~(\ref{angint}) is valid if
$\frac{M_{x}}{2\left(M_{y} - M_{x}\right)}>0$, which is true for a
phosphorene double layer. Note that for the anisotropic superfluid,
formed by paired fermions,
 the relation $n_{n}^{(xx)}(T) = n_{n}^{(yy)}(T)$ is also valid~\cite{Saslow}.

Under the assumption of the sound spectrum of collective excitations
using Eq.~(\ref{nnxyfinal}), implying $n_{n}^{(xx)}(T) =
n_{n}^{(yy)}(T)$, one obtains  from Eq.~(\ref{nnang}) the concentration of the normal
component $\tilde{n}_{n}(T)$ as
\begin{eqnarray}
\tilde{n}_{n}(T) =  n_{n}^{(xx)}(T) = n_{n}^{(yy)}(T) = \frac{2\zeta
(3)s(k_{B}T)^{3}\sqrt{M_{x}M_{y}}}{\pi (\hbar g n)^{2}}  .
\label{nnsound}
\end{eqnarray}
Therefore, in case of the sound-like spectrum of
collective excitations, the concentration of the superfluid
component $\tilde{n}_{s}(T)$ is given by
\begin{eqnarray}
\tilde{n}_{s}(T) =  n -  \frac{2  \zeta
(3)s(k_{B}T)^{3}\sqrt{M_{x}M_{y}}}{\pi (\hbar g n)^{2}}  .
\label{nssound}
\end{eqnarray}

It follows from Eqs.~(\ref{nnsound}) and~(\ref{nssound})  that for the sound-like spectrum of collective excitations, the concentrations of the
normal and superfluid components do not depend on an angle.

\begin{figure}[h]
\centering
\includegraphics[width=18.0cm]{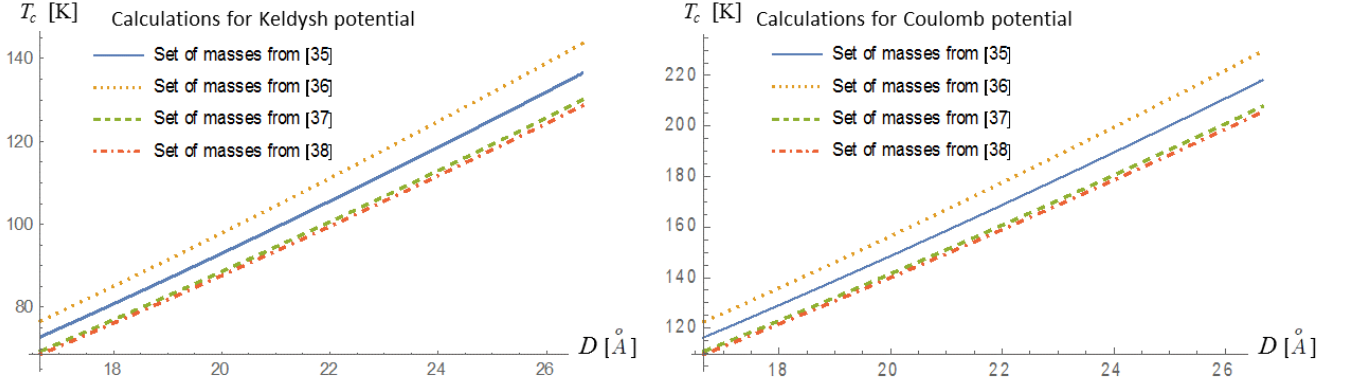}   \vspace{-2.7cm}
\caption{(Color online) The mean field critical temperature for
superfluidity $T_{c}$ for a phosphorene double layer as a function
of the interlayer separation $D$, assuming the sound-like spectrum
of collective excitations. (a) The interaction between the carriers
is described by the Keldysh potential. (b) The interaction between
the carriers is described by the Coulomb potential. The exciton
concentration is $n =  2 \times 10^{12} \ \mathrm{cm^{-2}}$.}
\label{Fig44}
\end{figure}

For the sound-like spectrum of collective excitations, the mean
field critical temperature $T_{c}$ can be obtained by substitution
Eq.~(\ref{nnsound}) into the condition $\tilde{n}_{n}(T_{c}) = n$ as
 following
\begin{eqnarray}
T_{c} = \left(\frac{\pi(\hbar g)^{2}}{2\zeta
(3)s\sqrt{M_{x}M_{y}}}\right)^{1/3}\frac{n}{k_{B}}  . \label{Tcso}
\end{eqnarray}
It follows from Eq.~(\ref{Tcso}) that under the assumption about the
sound-like spectrum of collective excitations, the mean field
critical temperature $T_{c}$ does not depend on an angle. The mean
field critical temperature of the superfluidity $T_{c}$ for the
Keldysh and Coulomb potentials for the
sound-like spectrum of collective excitations obtained by using Eq.~(\ref{Tcso}) as a
function of the interlayer separation $D$, is presented in
Fig.~\ref{Fig44}. The calculations are performed for the sets of effective electron and hole masses from Refs.~\onlinecite{Peng2014,Tran2014-2,Paez2014,Qiao2014}.
Comparing Fig.~\ref{Fig44}a with
Fig.~\ref{Fig44}b, one concludes that at the same parameters, the
critical temperature for the superfluidity $T_{c}(\Theta)$
 is much larger for the Coulomb
potential than for the Keldysh potential, because the sound velocity
for the Coulomb  potential is larger  than for the Kelsysh potential
due to the screening effects, implied by the Keldysh potential. However, for both potentials
the mean field critical temperature for superfluidity shows the similar depends on the electron and hole effective masses.

\medskip
\par
\begin{table}[b]
\caption{The critical temperatures under the    assumption about
 the sound-like spectrum of collective excitations
  for different sets of masses from Refs. \cite{Peng2014}, \cite%
{Tran2014-2}, \cite{Paez2014}, and \cite{Qiao2014}. The phosphorene
layers are separated by 7 layers of h-BN. $\mu _{0}$ and
$M_{x}M_{y}$ are expressed in units of free electron mass $m_{0}$
and $m_{0}^{2},$ respectively.}
\begin{center}
\begin{tabular}{ccccc}
\hline\hline Mass from Ref: & \cite{Peng2014} & \cite{Tran2014-2} &
\cite{Paez2014} & \cite{Qiao2014} \\ \hline
$\mu _{0},$ $\times $10$^{-2}m_{0}$ & 3.99 & 4.11 & 4.84 & 4.79 \\
Coulomb potential $T_{c},$ K & 182 & 192 & 174 & 172 \\
Keldysh potential $T_{c},$ K & 115 & 121 & 109 & 107 \\
$M_{x}M_{y},$ $\times m_{0}^{2}$ & 1.67 & 1.23 & 2.24 & 2.39 \\
\hline\hline
\end{tabular}
\end{center}
\label{tab2}
\end{table}

 As it is demonstrated  in Table~\ref{tab2},  the
critical temperature for the superfluidity $T_{c}$ decreases when
$M_{x}M_{y}$ increases. Therefore,  $T_{c}$ is sensitive to the
electron and hole  effective masses.

\medskip
\par

\begin{figure}[h]
\centering
\includegraphics[width=20.0cm]{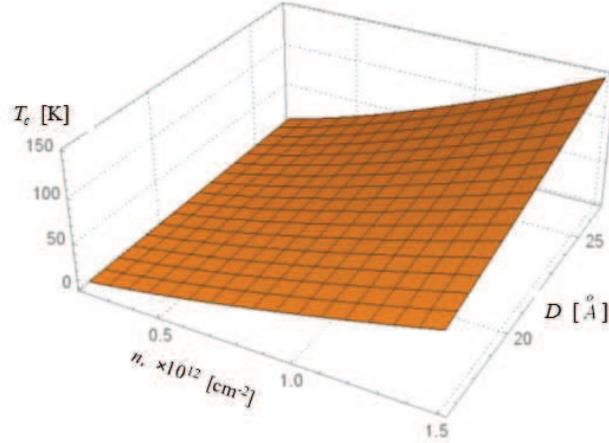}
\vspace{-3cm} \caption{(Color online) The critical temperature for
superfluidity $T_{c}$ for a phosphorene double layer as a function
of the exciton concentration $n$ and the interlayer separation $D$,
assuming the sound-like spectrum of collective excitations. The
calculations are performed for the Coulomb potential. The set of
masses is taken from Ref.~[\onlinecite{Paez2014}].} \label{Fig5}
\end{figure}

 Assuming the sound-like spectrum of collective
excitations, the mean field critical temperature of the
superfluidity $T_{c}$ obtained by using Eq.~(\ref{Tcso}) as a
function of the exciton concentration $n$ and the interlayer
separation $D$, is presented in Fig.~\ref{Fig5}. While the
calculations, presented in Fig.~\ref{Fig5}, were performed for the
Coulomb potential, one can obtain the similar behavior for the mean
field critical temperature of the superfluidity by employing the
Keldysh potential. According to Figs.~\ref{Fig44} and~\ref{Fig5},
the mean field critical temperature of the superfluidity $T_{c}$ is
an increasing function of the exciton concentration $n$ and the
interlayer separation $D$.

\subsection{Superfluidity when the  spectrum
of collective excitations is given by Eq.~(\ref{colsp})}

\begin{figure}[h]
\centering
\includegraphics[width=18.0cm]{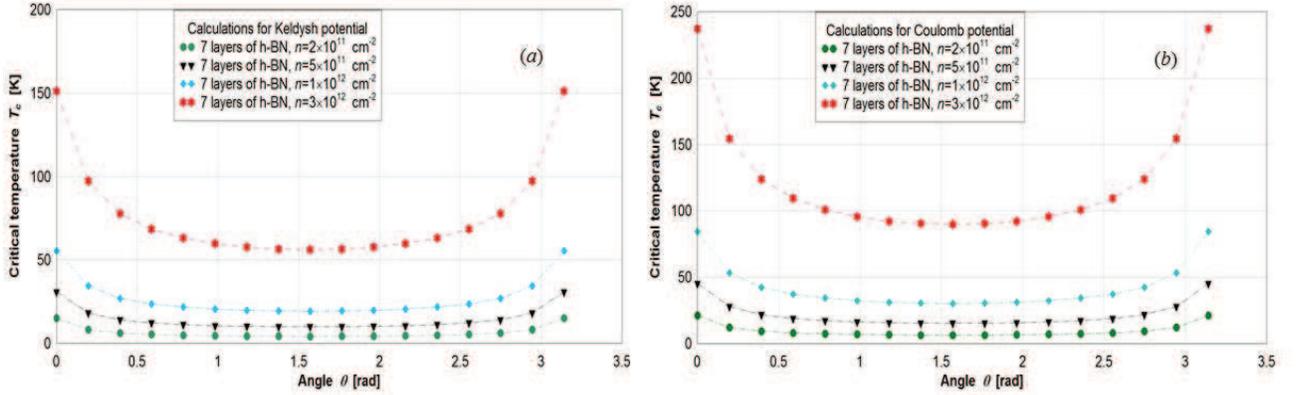}   \vspace{-2.4cm}
\caption{(Color online) The angular dependence of the critical
temperature for superfluidity $T_{c}(\Theta)$ for a phosphorene
double layer for  different exciton concentrations. (a) The
interaction between the carriers is described by the Keldysh
potential. (b) The interaction between the carriers is described by
the Coulomb potential. The number of $h$-BN monolayers between the
phosphorene monolayers is $N_{L} = 7$. The set of masses is taken from
Ref.~[\onlinecite{Paez2014}].} \label{Fig6}
\end{figure}

 Beyond the assumption of the sound-like spectrum,
substituting Eq.~(\ref{colsp}) for the spectrum of collective
excitations into Eq.~(\ref{nnxy}), and using Eq.~(\ref{nnang1}), we
obtain the mean field critical temperature of the superfluidity
$T_{c}(\Theta)$, by solving numerically Eq.~(\ref{Tc}). Since in
this case $n_{n}^{(xx)}(T) \neq n_{n}^{(yy)}(T)$, the mean field
critical temperature of the superfluidity $T_{c}(\Theta)$ is angular
dependent. The angular dependence of critical temperature
$T_{c}(\Theta)$ for the Keldysh and Coulomb potentials for different
exciton concentrations, calculated by
 solution of transcendental equation~(\ref{Tc}), is presented in
Fig.~\ref{Fig6}. According to Fig.~\ref{Fig6}, the mean field
critical temperature of the superfluidity $T_{c}(\Theta)$, is an
increasing function of the exciton concentration $n$. According to
Fig.~\ref{Fig6}, the critical critical temperature of the
superfluidity is maximal at $\Theta = 0$ and $\Theta = \pi$ and
minimal at $\Theta = \pi/2$.


As it follows from comparison of
Fig.~\ref{Fig6}a with Fig.~\ref{Fig6}b, at the same parameters, the
mean field critical temperature for the superfluidity
$T_{c}(\Theta)$   is greater when one considers the Coulomb
potential for the interaction between the charge carriers than for
the  Keldysh potential, because the sound velocity for the Coulomb
potential is greater than for the Keldysh potential due to the
screening effects, taken into account by the Keldysh potential.

It is interesting to mention that the ratio of the maximal critical
temperature $T_{c}^{(\mathrm{max})} = T_{c}(0)$ to the minimal
critical temperature $T_{c}^{\mathrm{(min)}} = T_{c}(\pi/2)$,
$T_{c}^{(\mathrm{max})}/T_{c}^{\mathrm{(min)}}$, in case of both the
Keldysh and Coulomb interactions between the charge carriers
decreases from $3.55$ to $2.69$ for the Keldysh potential, and from
$3.29$ to $2.64$ for the Coulomb potential, when the density of
exciton increases from $n = 2 \times 10^{11} \ \mathrm{cm^{-2}}$ to
$n = 3 \times 10^{12} \ \mathrm{cm^{-2}}$. One concludes that the
angular dependence of the mean field critical temperature $T_{c}$
decreases, when the exciton concentration increases.

At the fixed exciton concentration $n$, at the temperatures below
$T_{c}^{\mathrm{(min)}}$, exciton superfluidity exists at any
direction of exciton motion with any angle $\Theta$ relative to the
armchair direction, while  at the temperatures above
$T_{c}^{\mathrm{(max)}}$, exciton superfluidity is absent at any
direction of exciton motion with any angle $\Theta$.  At the fixed
exciton concentration $n$, at the temperatures in the range
$T_{c}^{\mathrm{(min)}} < T < T_{c}^{\mathrm{(max)}}$, exciton
superfluidity exists only for the directions of exciton motion with
the angles in the ranges $0< \Theta < \Theta_{c1}(T)$ and
$\Theta_{c2}(T)< \Theta < \pi$, while the superfluidity is absent
for the directions of exciton motion with the angles in the range
$\Theta_{c1}(T) < \Theta < \Theta_{c2}(T)$. The  critical angles of
superfluidity $\Theta_{c1}(T)$ and $\Theta_{c2}(T)$ correspond in
Fig.~\ref{Fig6} to the left and right crossing points of the
horizontal line at the temperature $T$ with the curve at the fixed
exciton concentration  $n$, respectively.

Let us mention that the critical temperature for the superfluidity
for a BCS-like fermionic superfluid with the anisotropic order
parameter does not depend on the direction of motion of Cooper pairs
because in this case $n_{n}^{(xx)}(T) =
n_{n}^{(yy)}(T)$~\cite{Saslow}.

 Let us mention that we chose to use
 the set of masses from Ref.~[\onlinecite{Paez2014}], because
this set results in higher exciton binding
energy. We used the number of $h$-BN monolayers between the
phosphorene monolayers $N_{L} = 7$ for Figs.~\ref{Fig4}
and~\ref{Fig6}, because higher $N_{L}$ corresponds to higher
interlayer separation $D$, which results in higher critical exciton
velocity of superfluidity equal to the sound velocity
$c_{S}(\Theta)$ and higher mean field critical temperature of the
superfluidity $T_{c}(\Theta)$.

\medskip
\par

According to Eq.~(\ref{nnang1}), the angular dependent concentration
of the normal component $\tilde{n}_{n}(\Theta, T)$ for $0 \leq
\Theta \leq \pi/2$ increases with $\Theta$ if $n_{n}^{(yy)}(T) >
n_{n}^{(xx)}(T)$ and decreases with $\Theta$ if $n_{n}^{(yy)}(T) <
n_{n}^{(xx)}(T)$. Therefore, at $n_{n}^{(yy)}(T) > n_{n}^{(xx)}(T)$
the superfluidity can exist only if $\Theta < \Theta_{c} (T)$, while
at $n_{n}^{(yy)}(T) < n_{n}^{(xx)}(T)$ the
superfluidity can exist only if $\Theta > \Theta_{c} (T)$, where $%
\Theta_{c}(T)$ is the critical angle of the occurrence of superfluidity.

\medskip
\par

For a chosen temperature, the critical angle $\Theta_{c}(T)$, which
corresponds to the occurrence of superfluidity, is given by the
condition
\begin{equation}
\tilde{n}_{n}(\Theta_{c} (T), T) = n \ .
\label{Thetac}
\end{equation}
Substituting  Eq.~(\ref{nnang1}) into Eq,~(\ref{Thetac}), one
obtains a closed form analytic expression for $\Theta _{c}(T)$ as
\begin{equation}
\Theta _{c}(T)=\frac{1}{2} \arccos \left[\frac{%
2n^{2}-\left(\left[n_{n}^{(xx)}(T)\right]^{2}+\left[n_{n}^{(yy)}(T)\right]^{2}\right)}{%
\left[n_{n}^{(xx)}(T)\right]^{2}-\left[n_{n}^{(yy)}(T)\right]^{2}}
\right] \ . \label{critang}
\end{equation}

\medskip
\par

\section{Proposed experiment to observe the angular dependent superfluidity
of dipolar excitons in a phosphorene  Double Layer}
\label{experiment}

\medskip
\par

The angular dependent superfluidity in a phosphorene  double layer
may be observed in electron-hole Coulomb drag experiments. The
Coulomb attraction between electrons and holes can introduce a
Coulomb drag that is a process in spatially separated conductors,
which enables a current to flow in one of the layers to induce a
voltage drop in the other one. In the case when the adjacent layer
is part of a closed electrical  circuit, an induced current flows.
The experimental observation of exciton condensation and perfect
Coulomb drag was claimed recently for  spatially separated electrons
and holes in GaAs/AlGaAs coupled quantum wells in the presence of
high magnetic field perpendicular to the quantum wells \cite{Nandi}.
A steady transport current of electrons driven through one quantum
well was accompanied by an equal current of holes in another. In
Ref.\  [\onlinecite{Pogrebinskii}],  the authors discussed the drag
of holes by electrons in a semiconductor-insulator-semiconductor
structure. The prediction was that for two conducting layers
separated by an insulator there will be a drag of carriers in one
layer due to the direct Coulomb attraction with the carriers in the
other layer. The Coulomb drag effect in the electron-hole double
layer BCS system was also analyzed in Refs.\  [\onlinecite{VM,JBL}].
If the external potential difference is applied to one of the
layers, it will produce  an electric current. The  current in an
adjacent layer will be initiated as a result of the correlations
between electrons and holes at temperatures below the critical one.
Consequently, the Coulomb drag effect was explored for semiconductor
coupled quantum wells in a number of theoretical and experimental
studies~\cite{Gramila,Sivan,Gramila2,Jauho,Zheng,Sirenko,Tso,Flensberg,Tanatar,EMN}.
The Coulomb drag effect in two coaxial nanotubes was studied in
Ref.\ [\onlinecite{BGK}].  The experimental and theoretical
achievements in Coulomb drag effect have been reviewed in Ref.\
[\onlinecite{Narrmp}].

\medskip
\par

We propose to study experimentally the angular dependent
superfluidity of dipolar excitons in a phosphorene  double layer by
applying a voltage difference for current flowing in one layer in a
chosen direction at a chosen angle $\Theta$ relative to the armchair
direction and measuring the drag current in the same direction in
another layer. This drag current in another layer in the same
direction as the current in the first layer will be initiated by the
electron-hole Coulomb drag effect due to electron-hole attraction.
The measurement of the drag current in an adjacent layer for a
certain direction with the corresponding $\Theta$ will indicate the
existence of superfluidity in this direction.  Due to the angular
dependence of the sound velocity, the critical exciton velocity for
superfluidity depends on an angle. Therefore, for certain exciton
velocities, there are the angle ranges, which correspond to the
superfluid exciton flow, and other angle ranges, which correspond to
the normal exciton flow. This can be applied as a working principal
for switchers, controlling the exciton flows in different directions
of exciton motion, caused by the Coulomb drag effect.

\medskip
\par

\section{Conclusions}
\label{conc}

In summary, the influence of the anisotropy of the dispersion
relation of dipolar excitons in  a double layer of black phosphorene
on the excitonic BEC and directional superfluidity has been
investigated. The analytical expressions for the single dipolar
exciton energy spectrum and wave functions have been derived. The
angle dependent spectrum of collective excitations and sound
velocity have been derived. It is predicted that a weakly
interacting gas of dipolar excitons in a double layer of black
phosphorus exhibits superfluidity at low temperatures due to the
dipole-dipole repulsion between the dipolar excitons. It is
concluded that the anisotropy of the energy band structure in a
black phosphorene causes the critical velocity of the superfluidity
to depend on the direction of motion of dipolar excitons. It is
demonstrated that the  dependence of the concentrations of the
normal and superfluid components and the mean field critical
temperatures for superfluidity on the direction of motion of dipolar
excitons occurs beyond the sound-like approximation for the spectrum
of collective excitations. Therefore, the directional superfluidity
of dipolar excitons in a phosphorene double layer is possible.
Moreover, the presented results, obtained for both Keldysh and
Coilomb potentials, describing the interactions between the charge
carriers, allow to study  the influence of the screening effects on
the dipolar exciton binding energy, exciton-exciton interaction, the
spectrum of collective excitations, and the critical temperature of
superfluidity for a weakly interacting Bose gas of dipolar excitons
in a phosphorene double layer. It is important to mention that the
binding energy of dipolar excitons, and mean field critical
temperature for superfluidity are sensitive to the electron and hole
effective masses. Besides, the possibilities of the experimental
observation of the  superfluidity for  various directions of motion
of excitons were briefly discussed.

\medskip
\par

 Our analytical and numerical results provide motivation for future experimental and theoretical
 investigations on excitonic BEC and superfluidity for double layer phosohorene.

\end{document}